\documentclass[twocolumn,amsmath,amssymb,prl,superscriptaddress,shortbibliography]{revtex4-1}

\setcitestyle{super}
\usepackage{bm}

\usepackage[final,letterspace=-40]{microtype}
\usepackage[colorlinks=true, linkcolor=blue, urlcolor=blue,  citecolor=blue, anchorcolor=blue]{hyperref}

\usepackage{mathpazo} 

\usepackage{graphicx}
\usepackage{dcolumn}
\usepackage{bm}
\usepackage{helvet}

\usepackage{caption}
\DeclareCaptionLabelFormat{plain}{Figure #2 $\boldsymbol |$ }

\begin{document}
\title{Interferometric control of the photon-number distribution}

\author{H. Esat Kondakci}
\email{esat@creol.ucf.edu}
\affiliation{CREOL, The College of Optics \& Photonics, University of Central Florida, Orlando, Florida 32816, USA}
\author{Alexander Szameit}
\affiliation{Institute for Physics, University of Rostock, 18059 Rostock, Germany}
\author{Ayman F. Abouraddy}
\author{Demetrios N. Christodoulides}
\author{Bahaa E. A. Saleh}
\affiliation{CREOL, The College of Optics \& Photonics, University of Central Florida, Orlando, Florida 32816, USA}

\begin{abstract}
We demonstrate deterministic control over the photon-number distribution by interfering two coherent beams within a disordered photonic lattice. By sweeping a relative phase between two equal-amplitude coherent fields with Poissonian statistics that excite adjacent sites in a lattice endowed with disorder-immune chiral symmetry, we measure an output photon-number distribution that changes periodically between super-thermal and sub-thermal photon statistics upon ensemble averaging. Thus, the photon-bunching level is controlled interferometrically at a fixed mean photon-number by gradually activating the excitation symmetry of the chiral-mode pairs with structured coherent illumination and \textit{without} modifying the disorder level of the random system itself.  
\end{abstract}

\small 
\maketitle

Optical interferometers implement deterministic field transformations that trace an interferogram by sweeping a phase -- which enables applications across all of optics and photonics \cite{Saleh2007}. The detected intensity is modulated, but the underlying photon number distribution $P_n$ does not change: the Poisson statistics associated with coherent light remain Poissonian, and the Bose-Einstein statistics that are the hallmark of thermal light remain Bose-Einstein. Introducing dynamical randomness in the interferometer \textit{can} change the photon statistics. An early example in neutron interferometry implemented a chopper in one arm of a two-path interferometer \cite{Rauch84PLA, Summhammer87PRA}, while an optical `stochastic interferometer' devised by De Martini \textit{et al.} changes the photon statistics by inserting a randomly varying optical element \cite{DeMartini89EPL, DeMartini1992, DeMartini1992a, DeMartini02JOSAB}. Along a different vein, coherent light traversing a random medium may gradually acquire chaotic behavior and under certain conditions may even attain Bose-Einstein statistics. This phenomenon has been observed in several systems, including weakly transmitting barriers embedded in a waveguide \cite{Kindermann2002}, quasi one-dimensional disordered samples \cite{Beenakker2000}, and optically dense slabs containing multiple scatterers \cite{Lodahl2005, Lodahl2006, Balog2006a, Smolka2011,Schlawin2012,Kropf2016}. In these approaches, changing the photon statistics is predicated on varying the system disorder level -- with the amount of fluctuations (photon bunching) typically proportional to the disorder level.

We present here a different strategy for tuning photon statistics in a random structure that does \textit{not} require modifying its disorder level. Instead, the ensemble-averaged $P_n$ is varied \textit{determinstically} by an \textit{external} element in the form of a relative phase between two input channels of a random network -- analogously to a traditional two-path interferometer. Reconstructing $P_n$ reveals deterministic interferometric tuning of photon-bunching across the super-thermal and sub-thermal regimes with \textit{fixed} mean photon-number $\bar{n}$. This effect has its origin in the smooth and deterministic breaking of the excitation symmetry in certain random lattices, which can be achieved by sculpting the lattice excitation. Our experiment makes use of a one-dimensional (1D) disordered photonic lattice of evanescently coupled waveguides with off-diagonal disorder \cite{Martin2011a}. Additionally, we measure two statistical quantities: the normalized second-order photon correlation $g^{(2)}$, which does not depend on $\bar{n}$, and Mandel's $Q$-parameter, which does\cite{Mandel1995}. In previous work \cite{Kondakci2016}, we measured the statistical parameter $g^{(2)}$. Although $g^{(2)}$ can be computed from the photon-number distribution, the converse is \textit{not} true. The counting distribution has features that are not captured by a single parameter, its normalized second-order moment $g^{(2)}$. The novelty in our work lies in revealing the tunability of the shape of the distribution itself, by demonstrating the transition between a Poisson distribution, to a Bose-Einstein-like distribution, and the emergence of other intermediate photon counting distributions.

\begin{figure*}[t]
\centering \includegraphics[scale=1.1]{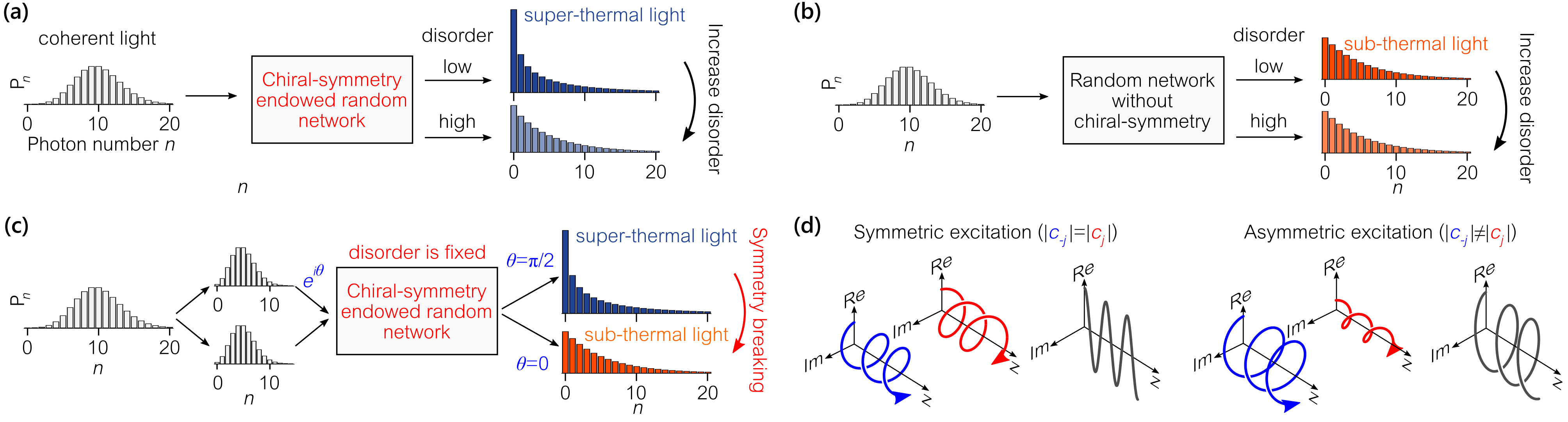} \caption{\label{Fig:Concept} \textbf{The concept of interferometric control over $P_n$.} \\
(a,b) Coherent light with Poissonian statistics is fed into a channel of a random network while varying the disorder level $\Delta$. The emerging light (a) spans the regime of super-thermal statistics when the lattice is endowed with chiral symmetry, or (b) spans the sub-thermal regime when the network lacks chiral symmetry. (c) Coherent light is split into two paths and a relative phase $\theta$ is introduced before being launched into a random network with chiral symmetry at fixed $\Delta$. Modulating $\theta$ breaks the mode-excitation symmetry and enables spanning the sub- and super-thermal regimes. (d) When the chiral symmetric eigenmode pairs in a lattice with of off-diagonal disorder (corresponding  phasors in a single lattice site depicted in blue and red) are activated with equal weights (symmetric excitation), the phasor sum (depicted in gray) takes on either real or imaginary values depending on the lattice site. This symmetry is absent  when the chiral symmetry is not activated (asymmetric excitation) or absent (diagonal disorder). Thus, the phasor sum is always complex (not only real or imaginary).}
\end{figure*}

The concept of interferometric control over $P_n$ is depicted in Fig.~\ref{Fig:Concept}. We start by considering coherent light with a Poissonian distribution $P_n = \bar{n}^{n}e^{-\bar{n}}/n!$, $n = 0,1,2,\cdots$, that is fed into a single channel of a disordered lattice consisting of an array of coupled optical elements (waveguides \cite{Christodoulides2003a, Schwartz2007a, Levi2011a}, resonators \cite{Topolancik2007, Mookherjea2008, Mookherjea2014}, or fiber loops \cite{Regensburger2011, Regensburger2012}). The output $P_n$ can be changed by varying the lattice disorder level $\Delta$ (to be defined below). Surprisingly, high-order coherences do \textit{not} decline while increasing $\Delta$ [Ref.~\onlinecite{Kondakci2015}]. If the lattice is endowed with disorder-immune `chiral symmetry' \cite{Dyson1953, Gade1993, Bocquet2003a}, a \textit{photonic thermalization gap} \cite{Kondakci2015b} emerges upon ensemble averaging: the regime of sub-thermal photon statistics is forbidden at \textit{any} disorder level, while super-thermal statistics are inaccessible to lattices lacking chiral symmetry \cite{Kondakci2015b}. We make use of the normalized second-order photon correlation $g^{(2)} = \langle n(n - 1)\rangle /\langle n\rangle^2$ ($\langle.\rangle$ denotes ensemble averaging over both the quantum state and the lattice disorder) as a scalar measure of randomness to delineate the sub-thermal $1 < g^{(2)} < 2$ and super-thermal $g^{(2)} > 2$ regimes \cite{Goodman2000}. Instead of a monotonic trend towards `thermalization' with increasing $\Delta$ in lattices characterized by chiral symmetry (such as those with \textit{off-diagonal} disorder \cite{Soukoulis1981a, Martin2011a}), $P_n$ exhibits \textit{super}-thermal statistics with a gradual decline towards thermal statistics upon increasing $\Delta$ [Fig.~\ref{Fig:Concept}(a)]. Such a lattice can thus tune the photon statistics in the \textit{super}-thermal regime. Alternatively, in lattices lacking chiral symmetry (such as those with \textit{diagonal} disorder \cite{Anderson1958a, Lahini2008a}), $P_n$ exhibits sub-thermal statistics with a gradual decline towards Poisson statistics upon increasing $\Delta$ [Fig.~\ref{Fig:Concept}(b)]. This lattice can thus tune the photon statistics in the \textit{sub}-thermal regime. 

To tune the photon statistics \textit{without} altering $\Delta$, we sculpt the excitation over \textit{multiple} lattice sites; e.g., by varying the relative phase between two sites. In traditional interferometry, a relative phase $\theta$ is introduced between two fields before they are superposed to trace an \textit{intensity interferogram} \cite{Saleh2007}. In the interferometric scheme introduced here, a relative phase $\theta$ is introduced between adjacent channels of a random network with chiral symmetry fed with coherent light. The two fields superpose \textit{within} the network and $P_n(\theta)$ measured in a single output channel reveals a \textit{deterministic} tuning of $P_n(\theta)$ between super-thermal and sub-thermal regimes while sweeping $\theta$ [Fig.~\ref{Fig:Concept}(c)].

The above-described effect stems from the properties of the eigenmodes and eigenvalues of a lattice endowed with chiral symmetry. To appreciate the underlying physics, consider a generic tight-binding lattice model with nearest-neighbor-only coupling. The complex field amplitude $E_x(z)$ at site $x$ after traveling along $z$ is described by a set of coupled differential equations,
\begin{equation}
-i\frac{\mathrm{d}E_x}{\mathrm{d}z}  = \beta_{x}E_x +C_{x,x+1}E_{x+1} +C_{x,x-1}E_{x-1},
\end{equation}
where $\beta_{x}$ is the wave number for site $x$ and $C_{x,x\pm1}$ the coupling coefficient between site $x$ and site $x\pm1$ -- all of which may be random variables when the lattice is disordered. We define a Hermitian coupling matrix or Hamiltonian $\mathbf{H}$ for the lattice where $\{\beta_{x}\}$ correspond to the diagonal elements and $\{C_{x,x\pm1}\}$ occupy the two next diagonals. The eigenvalues $\{b_{j}\}$ and eigenfunctions $\{\psi_{x}^{(j)}\}$ of $\mathbf{H}$ are defined through $\mathbf{H}\psi_{x}^{(j)} = (\bar{\beta}+b_{j})\psi_{x}^{(j)}$, where $\bar{\beta}$ is the average wave number. Since $\mathbf{H}$ is real and symmetric, $\{\psi_{x}^{(j)}\}$ are all real. If $\mathbf{H}$ can be recast into a \textit{block-diagonal} form after setting $\bar{\beta} = 0$ in the interaction picture, this indicates that the lattice is endowed with chiral symmetry. Lattices with \textit{off-diagonal} disorder ($\beta_{x} = \bar{\beta}$ and $C_{x,x\pm1}$ are random) satisfy chiral symmetry, whereas lattices featuring \textit{diagonal} disorder ($C_{x,x\pm1} = \bar{C}$ and $\beta_{x}$ are random) do not. Henceforth, we focus our attention on lattices with off-diagonal disorder. A consequence of chiral symmetry is that $b_{-j} = -b_{j}$ and $\psi_{x}^{(-j)} = (-1)^{x}\psi_{x}^{(j)}$. However, for the impact of this skew-symmetry to be manifested, the members of the skew-symmetric paired modes with indices $\pm j$ must be excited with equal weights, which we refer to as `activating' the chiral symmetry \cite{Kondakci2015b,Kondakci2016}. To visualize the activation of chiral symmetry, we associate a phasor with each mode at each lattice site. Then, in the rotating frame that is common to all modes, the eigenmodes appear in pairs whose members are equal magnitude counter-rotating phasors. This does not occur when the chiral symmetry is not activated or absent [Fig.~\ref{Fig:Concept}(d)]. 

\begin{figure*}[t]
\centering\includegraphics[scale=1.1]{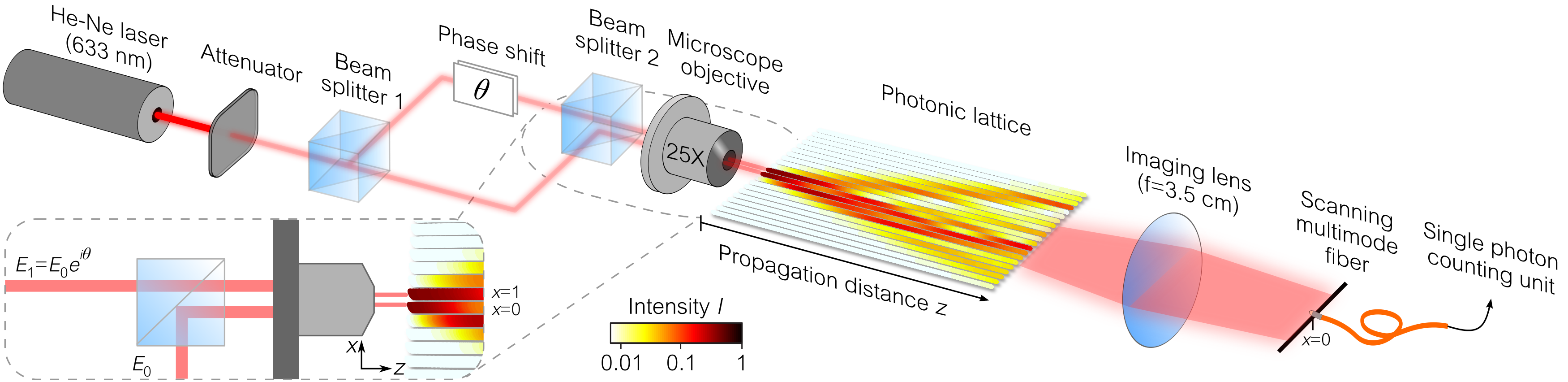}
\caption{\label{fig:Setup} \textbf{Experimental setup for interferometric control of the photon-number distribution.} \\
The input laser is attenuated, split into two paths (beam splitter 1), a relative phase $\theta$ is introduced, the two beams are brought together in parallel paths (beam splitter 2), and then imaged into two adjacent waveguides in the array sample. The intensity distribution along the photonic lattice is plotted (in logarithmic scale) for a single realization of lattice disorder. The inset shows the coupling scheme, where $E_0$ and $E_1 = E_0e^{i\theta}$ are the coherent field amplitudes coupled to waveguides $0$ and $1$, respectively. Furthermore, a beam splitter at the output directs light that is imaged to a CCD camera, which helps ensure coupling into two neighboring waveguides for each disorder realizations (the camera and beam splitter are omitted for clarity).}
\end{figure*}

This can be understood as follows. High-order coherences depend on the expected values of cross-correlations of the excited modal coefficients. In a disordered lattice, these typically all vanish except for the auto-correlation terms. However, in the presence of skew-symmetric modes in \textit{each} realization of the random ensemble, the cross-correlation of their excitation coefficients survive statistical averaging, thereby adding a contribution to photon bunching and resulting in super-thermal statistics. This can be seen at both low $\Delta$ via a perturbation analysis around the periodic lattice solutions, and high $\Delta$ where localization reduces the number of lattice modes coupled to the excitation and thus also reduces the photon bunching \cite{Kondakci2015b,Kondakci2016a}. One can also understand this behavior by examining the statistics of the field quadratures \cite{Kondakci_Topology}. Under symmetric excitation configurations, the average field is constrained to a single quadrature in the complex plane. As a result, the intensity distribution approaches to chi-squared distribution with one degree of freedom \cite{Papoulis1965} corresponding to modified Bose-Einstein photon number distribution \cite{Saleh1978}.

An input optical field $E_x(z = 0)$ can be analyzed in a basis of lattice eigenmodes, $E_x(0) = \sum_{j} c_{j}\psi_{x}^{(j)}$, where $c_{j} = \sum_{x}\psi_{x}^{(j)}E_x(0)$ is the amplitude of the $j^{\mathrm{th}}$ mode $\psi_{x}^{(j)}$. The field subsequently evolves after a distance $z$ into $E_x(z) = \sum_{j} c_{j}\psi_{x}^{(j)}e^{ib_{j}z}$. In the special case of a single-site excitation at the input $E_x(0) = \delta_{x,0}$, then $|c_{j}| = |c_{-j}| = |\psi_{0}^{(\pm j)}|$ for all $j$, so that chiral symmetry is activated. This is not necessarily the case for more general field excitations. For example, when two adjacent sites are excited equally $E_x(0) = \delta_{x,0}+\delta_{x,1}$, the modal coefficients are $c_{\pm j} = \psi_{0}^{(j)}\pm\psi_{1}^{(j)}$, and chiral symmetry is activated $|c_{j}| = |c_{-j}|$ only when the relative phase is $\theta = \pm\pi/2$. Gradually varying the phase $\theta$ for fixed relative amplitudes ($|A_{0}| = |A_{1}|$ in our experiment) tunes the chiral-symmetry breaking: maximal symmetry breaking at $\theta = 0$ or $\pi$, and symmetry activation at $\theta = \pm\pi/2$. 

The photonic lattice we utilized consists of an array of 101 identical 35-mm-long waveguides with nearest-neighbor-only evanescent coupling \cite{Meany2015}. The average separation between the waveguides is 17~$\mathrm{\mu}$m, resulting in an average coupling coefficient $\bar{C} \approx 1.71$~cm$^{-1}$ at a wavelength of 633~nm. The coupling coefficients are selected independently from a uniform probability distribution function with mean $\bar{C}$ and half-width $\Delta$ in units of $\bar{C}$; our sample has $\Delta \approx 0.6$. 

The optical arrangement used in demonstrating deterministic interferometric control over $P_n$ is illustrated in Fig.~\ref{fig:Setup}. A single-mode coherent beam from a He-Ne laser is attenuated by a neutral density filter and split into two paths via a beam splitter. A relative phase shift $\theta$ is introduced via a delay in one path varied in 20-nm steps. The two beams are then brought together by a second beam splitter in parallel but closely spaced paths, which are imaged to two neighboring waveguides. The output facet of the array is imaged with a magnification of $8\times$ via a lens (focal length $f = 3.5$ cm) to a plane in which we scan a multimode fiber whose core optimally couples light from an individual waveguide. The multimode fiber delivers the collected light to a single-photon-counting module, and the output photon-number distribution $P_n$ is reconstructed using three photon-detection time windows: 20, 40, and 60~$\mu$s. The input intensity level is reduced to low levels so that only a few photons are detected within these windows while minimizing the accidental arrivals of multiple photons. We generate different disorder realizations by moving the array along the $x$-direction and coupling into a new pair of waveguides \cite{Martin2011a}. We measure single realizations of $P_n(\theta)$ (averaged over $10^{4}$ shots of the detection window) at the central lattice site $x = 0$, and then average $P_n(\theta)$ over an ensemble of 15 disorder realization for each value of $\theta$ by shifting the input excitation site.

\begin{figure}[t]
\centering\includegraphics[scale=1.05]{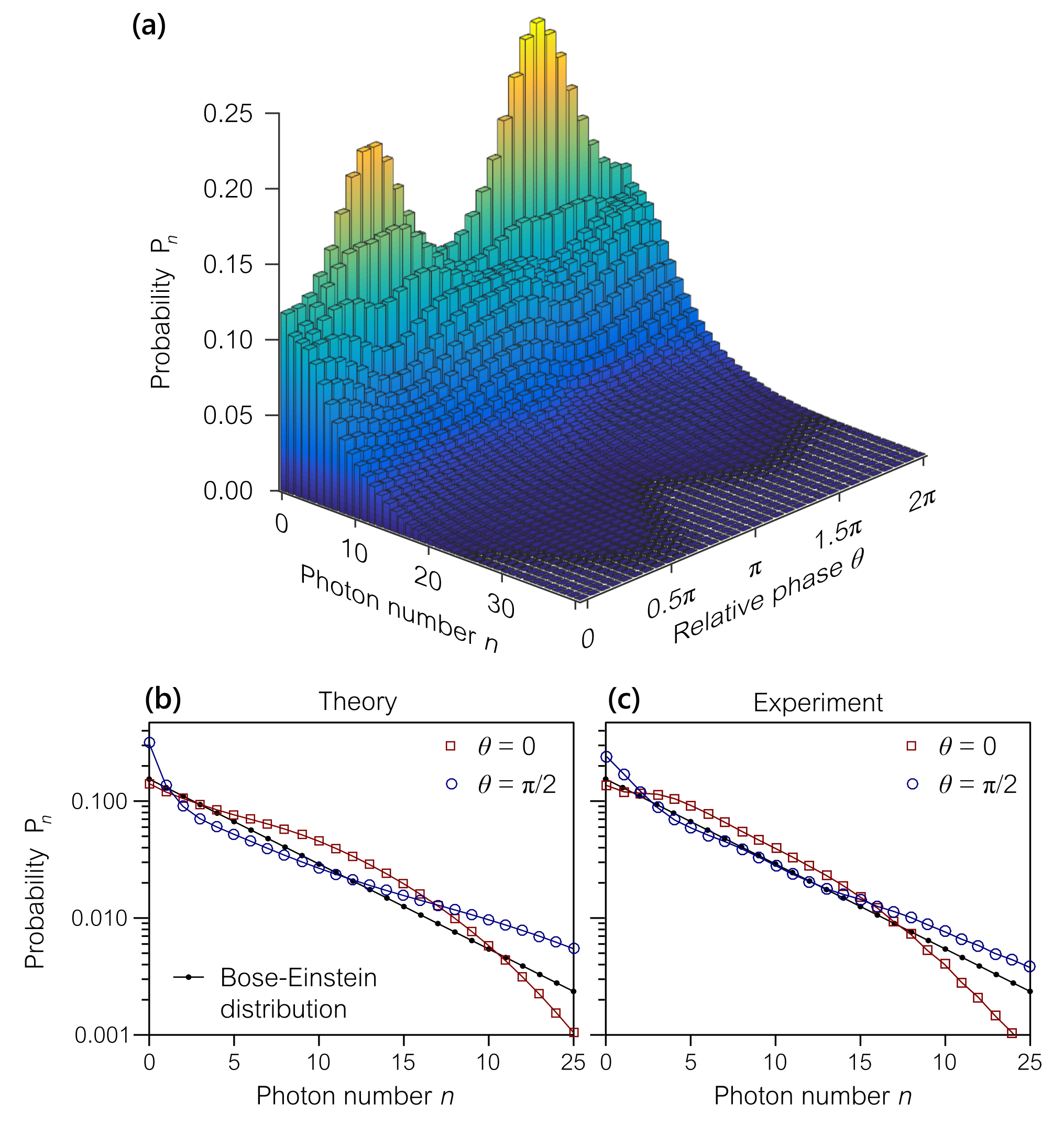}
\caption{\label{fig:PNSDistribution}\textbf{Deterministic interferometric tailoring of $P_n$.} \\ 
(a) The measured $P_n(\theta)$ while varying the phase $\theta$ between two coherent beams fed into adjacent sites of a disordered photonic lattice. The mean photon-number $\bar{n}(\theta) \approx 5.5$ is independent of $\theta$ (with a standard deviation $\approx 0.5$). (b) Simulated $P_n$ corresponding to $\theta = 0$ and $\pi/2$ obtained for $\bar{n} = 5.5$ and utilizing the physical parameters of the lattice. (c) Measured $P_n$, extracted from (a) and corresponding to the simulations in (b). (a,b) Black-dotted line is Bose-Einstein distribution for $\bar{n}=5.5$}
\end{figure}

We present in Fig.~\ref{fig:PNSDistribution} our measurements confirming the deterministic interferometric tuning of $P_n(\theta)$ in the excitation waveguide ($x = 0$) while varying $\theta$. As $\theta$ is swept, $P_n(\theta)$ varies periodically (period $\pi$) between sub-thermal to super-thermal statistics [Fig.~\ref{fig:PNSDistribution}(a)]. The measured mean photon-number $\bar{n}(\theta)$ does not vary with $\theta$ ($\bar{n} \approx 5.5$ with a standard deviation $\approx 0.5$). Although $\bar{n}$ remains constant, the photon-number distribution itself varies with $\theta$, achieving maximal bunching when the chiral symmetry is fully activated $\theta  = \pi/2,3\pi/2$, and minimal bunching at $\theta = 0,\pi$ when chiral symmetry is dormant. To better examine the salient changes in the photon statistics with $\theta$, we plot in Fig.~\ref{fig:PNSDistribution}(b,c) $P_n(\theta)$ for the extrema at $\theta = 0$ and $\pi/2$. The increase in the probability of higher photon numbers at the distribution tail when $\theta = \pi/2$ as compared to that when $\theta = 0$ is a clear signature of photon bunching. In comparison, we also plot the Bose-Einstein distribution for thermal light  with the experimentally determined $\bar{n} \approx 5.5$. At high $n$, the symmetric-excitation statistics exhibit higher probabilities than the Bose-Einstein distribution, whereas the broken-symmetric excitation has lower. Our experimental data [Fig.~\ref{fig:PNSDistribution}(c)] is in excellent agreement with the simulations [Fig.~\ref{fig:PNSDistribution}(b)].

\begin{figure}[t]
\centering\includegraphics[scale=1.05]{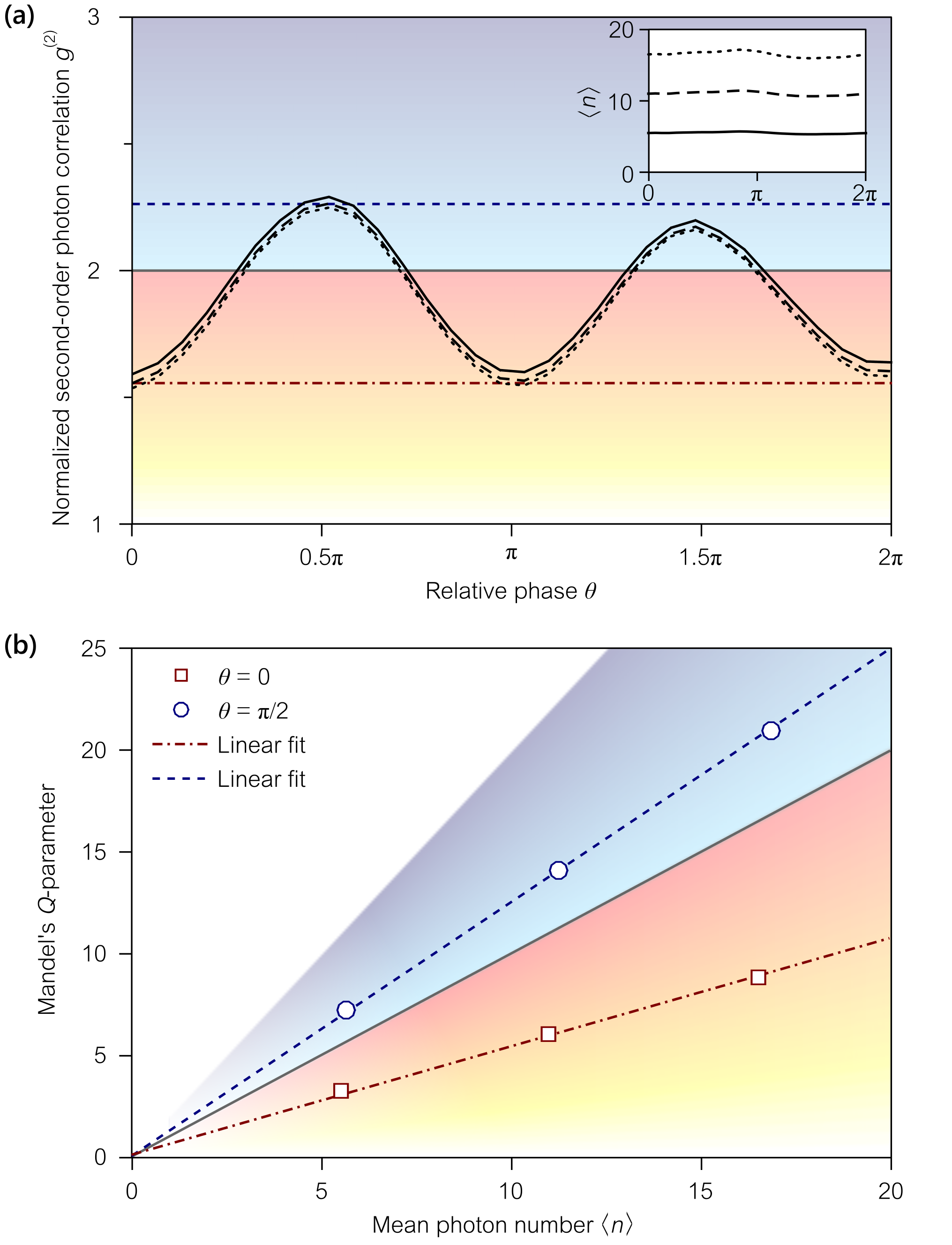} \caption{\label{fig:Q}\textbf{Deterministic interferometric control over  $g^{(2)}(\theta)$ and Mandel's $Q$-parameter.} \\ 
(a) $g^{(2)}$ as a function of $\theta$ obtained from the measured $P_n(\theta)$ for photon-counting windows of 20, 40, and 60~$\mu$s. The $g^{(2)}$-interferograms show no dependence on $\bar{n}$. The inset shows $\bar{n}(\theta)$ for the different photon-counting windows: $\bar{n}$ increases with the detection window but is independent of $\theta$. (b) Mandel's $Q$-parameter as a function of $\bar{n}$ at $\theta = 0$ (minimally activated chiral mode pairs) and $\theta = \pi/2$ (maximally activated chiral mode pairs) obtained from the measured $P_n$. The data points correspond to photon-counting windows in (a). The blue dashed line and the red dot-dashed lines are linear fits that fall in the super-thermal (blue-black color scheme) and sub-thermal (white-red color scheme) regimes, respectively. The border between the two color schemes demarcated by a solid gray line corresponds to Bose-Einstein statistics (thermal light). The dashed inclined lines correspond to horizontal dashed lines in (a).} 
\end{figure}

Further analysis of the measured distributions $P_n(\theta)$ helps bring about the changes that take place in the photon statistics. First, we examine a quantity extracted from $P_n(\theta)$ that does \textit{not} depend on $\bar{n}$ for the field considered here: the normalized second-order photon correlation function $g^{(2)}(\theta)$ [Ref.~\cite{Mandel1995}]. We plot in Fig.~\ref{fig:Q}(a) $g^{(2)}(\theta)$ and note clearly that it varies sinusoidally with $\theta$, between the sub-thermal and super-thermal regimes. The statistics are tuned from super-thermal ($g^{(2)} > 2$) to sub-thermal ($g^{(2)} < 2$), all while maintaining $\bar{n}$ fixed with $\theta$ [Fig.~\ref{fig:Q}(a), inset]. Although $\bar{n}$ changes according to the photon-counting window ($\bar{n} \approx 5.5, 11, 16.5$ for counting windows of $20, 40, 60$~$\mu$s, respectively), the three interferometric traces of $g^{(2)}(\theta)$ are indistinguishable.  

Next we consider a quantity that \textit{does} indeed depend on $\bar{n}$: Mandel's $Q$-parameter, $Q = \mathrm{Var}(n)/\langle n\rangle -1 = \bar{n}(g^{(2)}-1)$, which is thus linear in $\bar{n}$ for the field considered here \cite{Mandel1995}. Varying $\theta$ at a fixed $\bar{n}$ modulates $Q(\bar{n};\theta)$ between two limits identified by the dashed inclined lines in Fig.~\ref{fig:Q}(b); we identify in Fig.~\ref{fig:Q}(b) only the values corresponding to $\theta = 0$ and $\pi/2$, which are the extrema of this oscillation between the super-thermal and sub-thermal regimes. Increasing $\bar{n}$ (longer counting windows) leads to a linear growth in the two limits of $Q$ modulation.

In conclusion, we have demonstrated that the photon-number distribution $P_n$ -- and hence any photon statistic such as $g^{(2)}$ or Mandel's $Q$-parameter -- can be tuned deterministically by varying a relative phase between two equal-amplitude beams launched into adjacent sites of a disordered photonic lattice. Such interferometric control over $P_n$ is possible by judicious excitation-symmetry breaking of the skew-symmetric chiral mode pairs in disordered lattices that exhibit such symmetries.

Coupling light into a single site is a highly symmetric configuration. By sculpting the complex spatial distribution of the excitation over multiple sites, one may gradually break the excitation symmetry, which facilitates a smooth transition across the edge of the thermalization gap. This interferometric control strategy can be generalized in multiple ways. One may vary the relative amplitude of the fields exciting the two lattice sites while maintaining a fixed relative phase, or one may excite an extended section of the lattice while alternating the relative phases between adjacent sites. The latter approach has the advantage of producing the same phase-tunable $P_n$ with fixed $\bar{n}$ across the lattice.

Finally, we note that multiple experiments have been reported in which non-classical light is coupled into ordered or disordered photonic lattices\cite{Bromberg2009a,Lahini2011a,DiGiuseppe2013,Gilead2015}. In these reports, only small photon numbers are involved (usually two) in either Fock or entangled photon states, and while the output states different from those at the input, there have been no studies in the nature of the underlying photon statistics. Our work extends these studies into photon distributions with higher mean photon-number and demonstrates unambiguously interferometric control over $P_n$ itself.

A.F.A. was supported in part by the U.S. Office of Naval Research (ONR) under contract N00014-14-1-0260.

\bibliography{library_short,books,extras}
\end{document}